\documentclass[twocolumn,showpacs,preprintnumbers,amsmath,amssymb,pre]{revtex4-1}
\usepackage{color}

\usepackage{graphicx}
\usepackage{dcolumn}
\usepackage{bm}
\usepackage{subfigure}
\usepackage{amssymb}
\usepackage{multirow}
\graphicspath{{plots/}}

\begin{document}

\title{Comment on ``Discussion on ‘Novel attractive force
between ions in quantum plasmas -- failure of simulations based on
a density functional approach’''}

\author{M. Bonitz, E. Pehlke, and T. Schoof}%
\affiliation{%
    Christian-Albrechts-Universit\"at zu Kiel, 
    Institut f\"ur Theoretische Physik und Astrophysik, 
    Leibnizstra\ss{}e 15, 24098 Kiel, Germany
}

\date{\today}

\begin{abstract}
In a recent article [P.K. Shukla, B. Eliasson and M. Akbari-Moghanjoughi, Physica Scripta {\bf 87}, 018202 (2013)] the authors criticized 
our analysis of the screened proton potential in dense hydrogen that was based on {\em ab initio} density functional theory (DFT) simulations [M. Bonitz, E. Pehlke, and T. Schoof, Phys. Rev. E  {\bf 87}, 037102 (2013)]. 
In particular, they attributed the absence of the Shukla-Eliasson attractive force between protons in the DFT simulations to a failure of DFT. Here 
we discuss in detail their arguments and show that their conclusions are incorrect.
\end{abstract}

\maketitle

\section{Introduction}\label{s:intro}
Dense plasmas exhibiting quantum degeneracy of the electrons and Coulomb correlation effects are presently of high interest in many fields including condensed matter, astrophysics, warm dense matter and laser plasmas, e.g. \cite{horing_cpp11,kremp-etal.pre99,kremp-book,bonitz_pop08,militzer12}.
Even for the simplest plasma system -- hydrogen -- interesting questions remain partly open, including details of the phase diagram, e.g. \cite{ashcroft01} or the optical and transport properties under high compression, e.g. \cite{redmer}. 
Therefore, in recent years substantial efforts have been made to advance the theoretical description of quantum plasmas. There are two main lines of research: The first are analytical approaches such as quantum kinetic theory and nonequilibrium Greens functions methods, e.g. \cite{kremp-book,bonitz-book,bonitz-aip12} and linear response theory, e.g. \cite{roepke}. The second line comprises very successul developments of first-principle computer simulations -- including quantum Monte Carlo, e.g. \cite{ceperley,filinov00,schoof11}, quantum molecular dynamics, e.g. \cite{desiarlais}, density functional theory and combinations thereof, e.g. \cite{militzer12}. These methods are much more challenging than the analogous numerical approaches for classical plasmas due to the necessity to accurately include quantum and exchange effects and are still subject to active developments. We also mention efficient semiclassical approaches that use effective quatum pair potentials in classical simulations, e.g. \cite{afilinov_jpa03, afilinov_pre04} and references therein or classical--quantum mappings \cite{dharma,dufty13,dufty13_2}.

For an understanding of the above mentioned properties and many-particle effects of dense quantum plasmas a key question is how the microscopic Coulomb pair interaction between the ions is modified by the surrounding plasma. These modifications are due to screening, quantum and spin effects and lead to a replacement of the familiar Coulomb potential, $\phi^i(r)=Q/r$, of an ion observed {\em in vacuo}, by an effective potential. At weak non-ideality (i.e. weak interaction effects, weak coupling, see below) this gives rise to an isotropic Yukawa-type potential, 
$\phi^i_s(r)=\frac{Q}{r} e^{-r/l_s}$, where $l_s$ is the screening length. While in the limit of a classical high-temperature plasma $l_s$ is given by the Debye radius, in a dense quantum plasma, it is given by the Thomas-Fermi length $L_{\rm TF}$. More general screened potentials are successfully computed using linear response theory to obtain 
the dynamic dielectric function, $\epsilon^l({\bf k},\omega)$ (longitudinal density response), giving rise to \cite{diel_function,diel_tensor}
\begin{equation}
 \phi^i_s({\bf r}) = \frac{Q}{2\pi^2} \int d^3k \, \frac{e^{i{\bf k \cdot r}}}{k^2 \epsilon^l({\bf k}, {\bf k \cdot v})}.
 \label{eq:phi_general}
\end{equation}
The potential (\ref{eq:phi_general}) typically decays slower than the exponential Yukawa potential but faster than the Coulomb potential. 

For dense quantum plasmas expressions for the longitudinal dielectric function were first derived 60 years ago by Klimontovich and Silin \cite{klimontovich} and Bohm and Pines \cite{bohm_pines} and correspond to the mean field approximation (quantum Vlasov or Hartree approximation or RPA), in the condensed matter community this dielectric function is typically called the Lindhard function. Much more involved is the dielectric function for a correlated quantum plasma. Here substantial theoretical efforts are still under way, e.g. \cite{kwong_etal.prl00} and references therein.

In quantum plasmas the potential (\ref{eq:phi_general}) is largely determined by the degenerate electrons surrounding the ion. While the range of this potential is typically reduced by many-particle and quantum effects compared to the Coulomb case, the associated force between two ions -- as a rule -- retains its repulsive character.
However, in quantum plasmas in thermodynamic equilibrium \cite{wakes} two effects are known which can make this force attractive: i) The formation of bound states (such as hydrogen molecules or molecular ions). This evidently corresponds to a net attractive force between the constituents (hydrogen atoms, protons) whereby electronic spin effects play a crucial role. ii) Oscillatory potentials with (very shallow) negative parts (Friedel oscillations) emerge in a strongly degenerate Fermi gas as a consequence of the step character of the zero-temperature momentum distribution, e.g.\ \cite{kittel} and are well confirmed experimentally, e.g. \cite{friedel_exp}.

In a recent letter \cite{shukla_prl} Shukla and Eliasson (SE) claimed to have observed a ``novel attractive force'' between protons in a dense quantum hydrogen plasma at zero temperature. This claim was disputed by the present authors \cite{bonitz_pre_comm} on the basis of {\em ab initio} density functional theory (DFT) simulations and general considerations. In Ref. \cite{bonitz_pre_comm} we showed that the prediction by SE is incorrect and is caused by the use of linearized quantum hydrodynamic theory (QHD) beyond the limits of its validity. Nevertheless, the idea of the SE attractive potential has been used in a number of recent papers, despite its contradiction to {\em ab initio} results which underlines the importance of additional clarification.

One reason for this is that QHD (both, the nonlinear and linear [denoted ``LQHD'']versions) has become quite popular in recent years because it promises a semianalytical approach to quantum plasmas that avoids the complicated treatment of quantum and exchange effects in the previous theories and simulations (see above).
A number of authors \cite{vranies,vranies_re,vladimirov} raised serious concerns against its uncritical use that frequently disregards the applicability range of QHD. Our analysis presented in Refs. \cite{bonitz_pre_comm,bonitz_pre_reply} answers just the question about these limitations and is in agreement with the works \cite{vranies,vladimirov}.

The proponents of the SE-potential responded to this critique by a number of articles \cite{shukla_vranies,shukla_answer}. Another such response \cite{shukla_pscr} appeared recently in Physica Scripta where Shukla, Eliasson and Akbari-Moghanjoughi (SEA) claimed that our DFT results \cite{bonitz_pre_comm} are ``incomplete'' and lack ``the energy associated with the quantum recoil effect in the DFT Hamiltonian'' (citation from Ref.~\cite{shukla_pscr}, p. 3). According to SEA this is the reason why the Shukla-Eliasson attractive force is absent in the DFT simulations. This statement is wrong. The goal of the present paper is to respond to that article and to provide the necessary clarification.

This paper is organized as follows.
In Sec.~\ref{s:lqhd} we briefly recall the basic parameters of quantum plasmas and the parameters introduced by SEA (the basic formulas of QHD and LQHD used in Ref. \cite{shukla_prl} are given in the Appendix). In Sec.~\ref{s:dispute} we recall the recent dispute about the Shukla-Eliasson attractive force. Finally, analyzing, in Sec.~\ref{s:arguments} in detail the arguments of Shukla, Eliasson and 
Akbari-Moghanjoughi presented in Ref.~\cite{shukla_pscr}, we show that their conclusions are incorrect and that our critical analysis of the SE attractive force \cite{bonitz_pre_comm,bonitz_pre_reply} remains fully valid.

\section{Parameters of quantum plasmas and Linearized Quantum Hydrodynamics (LQHD)}\label{s:lqhd}
\begin{figure}[h]
\includegraphics[scale=0.85]{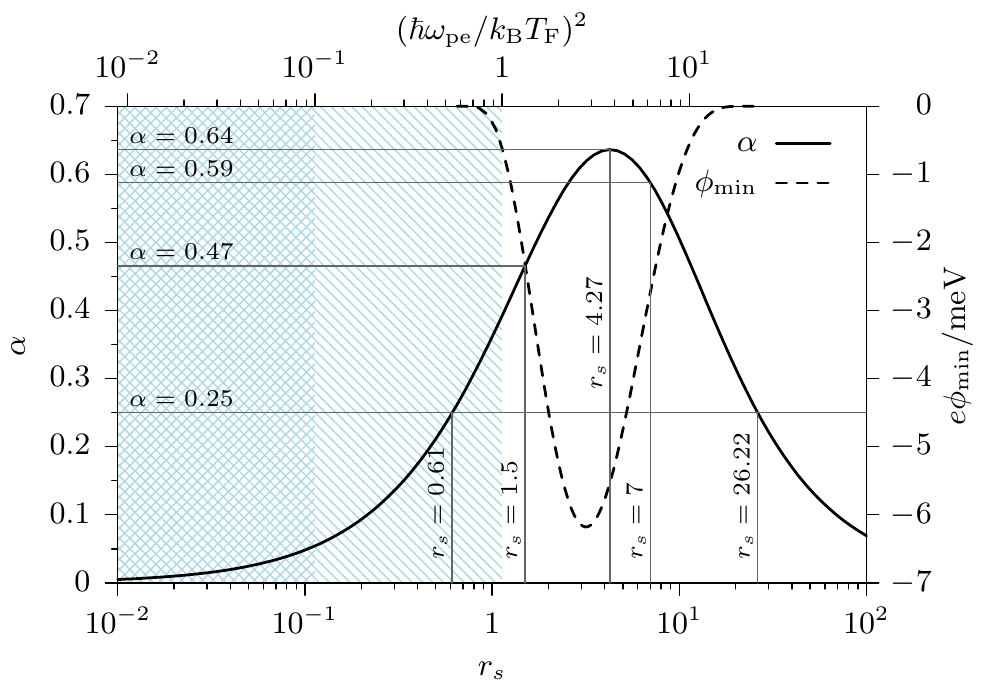}
\caption{
The Shukla-Eliasson parameter $\alpha$ (full line, left axis) versus Brueckner parameter $r_s$ (bottom axis) and $\Gamma_q$ (upper axis). The potential 
of a proton 
%, Eq.\ (\ref{eq:phi_general}), 
derived by Shukla and Eliasson from LQHD becomes attractive for $\alpha > 0.25$, corresponding to 
$26.22 \ge r_s \ge 0.61$, at zero temperature. The depth of the SE potential minimum is shown by the dashed curve (right axis).
The two shaded areas denote the range of moderate (weak) coupling given by $(\hbar \omega_{\rm pe}/k_{\rm B} T_{\rm F})^2 < 1$ ($<0.1$). } 
\label{fig:alpha}
\end{figure}

Dense quantum plasmas in condensed matter, astrophysics or warm dense matter contain (at least) two charged components. Ions, due their large mass, are typically classical and moderately or strongly coupled. Electrons, in contrast, are often partially or strongly degenerate but weakly correlated \cite{bonitz_pop08}. For the discussion of Ref. \cite{shukla_pscr} we only need to consider the electrons whereas ions appear as test charges screened by the electrons. Electrons are characterized by the following {\em length scales}, e.g. \cite{bonitz_pop08}: the mean interparticle distance ${\bar r}$, the Bohr radius, $a_B$,
the Thomas-Fermi wavelength, $\lambda_F=2\pi/k_F$, 
\begin{eqnarray}
 {\bar r} &=& [3/(4\pi n)]^{1/3}, 
\nonumber\\
a_B &=& 4\pi\epsilon_0\hbar^2/(me^2),
\nonumber\\
\lambda_F &=& 2\pi (3\pi^2 n)^{-1/3},
\label{eq:lambda_f}\\
\lambda_B &=& h/\sqrt{2\pi mk_BT},
\label{eq:lambda_b}
\end{eqnarray}
where $n$ denotes the mean electron number density.
At finite temperature, another important scale is the DeBroglie wave length, $\lambda_B$, Eq.~(\ref{eq:lambda_b}). Similarly, important {\em energy scales} are the hydrogen binding energy $E_R$ (referring to the atomic ground state), the plasmon energy $\hbar \omega_p$ with the plasma frequency $\omega_p=(ne^2/\varepsilon_0 m)^{1/2}$, the Fermi energy, $E_F$ and the thermal energy, $k_BT$,
\begin{eqnarray}
 E_R &=& e^2/(4\pi\epsilon_0 2a_B),
\nonumber\\
E_F &=& (\hbar k_F)^2/2m,
\label{eq:e_f}
\\
\langle U \rangle &=& e^2/(4\pi\epsilon_0{\bar r}),
\end{eqnarray}
where $\langle U \rangle$ characterizes the mean interaction energy of the electrons.
The properties of the electron gas are then characterized by several dimensionless parameters: 
the quantum degeneracy parameter, $\chi$,  and the coupling parameter that measures the interaction strength relative to the mean kinetic energy. Here the common choice in atomic physics and condensed matter theory is the Brueckner parameter, $r_s$. A very similar quantum coupling parameter 
is $\Gamma_q$ involving the plasmon energy, e.g. \cite{vladimirov}. The two coupling parameters have very similar values, as can be seen in Fig.~\ref{fig:alpha}.
\begin{eqnarray}
 \chi &=& n\lambda_B^3,
\label{eq:chi}\\
r_s &=& {\bar r}/a_B,
\label{eq:rs}\\
\Gamma_q &=& (\hbar \omega_p/E_F)^2.
\nonumber
\end{eqnarray}
Note that these coupling parameters are relevant for the ground state of the electron gas, $T=0$, which is considered by QHD. Furthermore,  
at $T=0$, obviously, the electrons are completely degenerate, $\chi \to \infty$

SEA in their paper \cite{shukla_pscr} use different parameters: a modified Brueckner parameter which we distinguish by the notation $r_s^{SEA}$. Further, they define a characteristic wave number $k_s$ and another coupling parameter $\alpha$. 
\begin{eqnarray}
 r_s^{SEA} &=& r_0/a_B,
\label{eq:rs_sea}\\
k_s &=& \omega_{pe}/\sqrt{v_*^2/3+v_{ex}^2},
\nonumber\\
\alpha &=& \hbar^2\omega_{pe}^2/4m_*^2(v_*^2/3+v_{ex}^2)^2,
\label{eq:alpha}
\end{eqnarray}
where $r_0=n^{-1/3}$, and the definitions of the velocities $v_*$ and  $v_{ex}$ are given in the Appendix. 
Note that the parameter $\alpha$ is, in fact, not a measure of the coupling strength since it depends non-monotonically upon the electron density (and on $r_s$ and $\Gamma_q$). This is illustrated in Fig. \ref{fig:alpha}.

\section{The dispute around the Shukla-Eliasson attractive potential}\label{s:dispute}

In a recent letter \cite{shukla_prl} Shukla and Eliasson (SE) applied the linearized QHD theory to the problem of the screened potential (\ref{eq:phi_general}) of a proton in a quantum degenerate electron gas at $T=0$. The equations of LQHD used by SE in Ref.~\cite{shukla_prl} are summarized in the Appendix where we also took into account the corrections published by SE in two Errata to the original paper.
SE found that this potential of a proton in dense hydrogen has a negative minimum for $\alpha>0.25$. In this case they obtained an explicit expression for the potential given by Eq.~(\ref{eqap:cos_pot}). Based on the existence of this negative minimum SE claimed the discovery of 
 a ``novel attractive force'' between protons in a dense hydrogen plasma and, further, the formation of bound states, a proton lattice, phase transitions and a critical point \cite{shukla_prl}. 

In Ref.~\cite{bonitz_pre_comm} the present authors studied the predictions of Shukla and Eliasson put forward in Ref.~\cite{shukla_prl}. We started by relating the condition $\alpha>0.25$ to the parameters of hydrogen and observed that it corresponds to a density range $0.61 \le r_s \le 26.22$, cf. Fig.~\ref{fig:alpha}. This range 
includes moderate electronic correlations (observed for $0.1 \lesssim r_s \lesssim 1$) and strong correlations ($r_s \gtrsim 1$). We then analyzed the properties of the potential and compared it to {\em ab initio} DFT results for the proton potential in dense hydrogen. The conclusions of Ref.~\cite{bonitz_pre_comm} are summarized as follows:
\begin{enumerate}
 \item The maximum depth of the SE potential is about $6$ meV which is negligible compared to relevant energy scales such as the binding energy of hydrogenic bound states (which is of the order of $E_R$) and the temperature of dense hydrogen plasmas in the laboratory or in astrophysical systems (which is at least several $eV$). 
 \item The distance of this minimum from a proton is $3\dots 10$ times larger than the mean interparticle distance in the relevant density range. Therefore, even if this minimum would be a real effect, protons in hydrogen could never arrange at such a distance from each other and, in particular, not form a lattice. Note that at the mean interparticle distance the SE potential is strongly repulsive \cite{shukla_prl}.
 \item The SE potential does not describe hydrogen at low density, $r_s>1$ correctly. One manifestation of this is that it fails to reproduce the formation of bound states ($r_s \gtrsim 1.5$).
 \item The SE potential does not describe hydrogen at high density correctly ($r_s \ll 1$), where it fails to reproduce the formation of Friedel oscillations.
 \item In the whole range of densities, the SE potential is qualitatively different from the DFT result and shows the opposite trends when the density is increased. 
\end{enumerate}
In contrast to the SE potential, the DFT was shown to reproduce the hydrogen molecule with the correct binding length and the Friedel oscillations, but not to exhibit any other attractive potential.
From this analysis, and based on the extensive available experience about the accuracy of DFT, we concluded in \cite{bonitz_pre_comm} that the LQHD results of Ref. \cite{shukla_prl} are incorrect and that there is no basis for the above mentioned predictions of SE for a dense hydrogen plasma. As the reason for the incorrect proton potential we identified that SE used LQHD outside its range of validity,  i.e. beyond the limit of an almost ideal Fermi gas. While in Ref. \cite{bonitz_pre_comm} we gave a conservative estimate for the validity range: $r_s<1$, in fact, weak nonideality of the electrons at zero temperature require an even stricter condition, i.e. $r_s \ll 1$ or, equivalently, $\Gamma_q \ll 1$. The density ranges  corresponding corresponding to weak ($r_s \lesssim 0.1$) and moderate ($0.1 \lesssim r_s \lesssim 1$) coupling are indicated in Fig.~\ref{fig:alpha} by the two shaded areas. We will come back to the dispute about the validit range of LQHD in Sec.~\ref{s:arguments}, cf. item ii. Notice, however, that even the ideal Fermi gas ($r_s \to 0$) is not described correctly by LQHD because this model neglects kinetic effects that are important to reproduce Friedel oscillations \cite{bonitz_pre_comm, vladimirov}. 

In response to our analysis in Ref. \cite{bonitz_pre_comm} Shukla, Eliasson and Akbari-Moghanjoughi (SEA) published a paper \cite{shukla_answer} where they rejected our conclusions. As the origin of the discrepancy between LQHD and DFT they put forward a failure of DFT. Furthermore, SEA objected against the comparison of LQHD and DFT for strong coupling, $r_s > 1$, although just such LQHD-data were presented in the original Letter \cite{shukla_prl}. The present authors had the opportunity to respond to the statements of SEA \cite{shukla_answer} in a Reply \cite{bonitz_pre_reply} where a comparison of the accuracy and applicability ranges of DFT and QHD were given. This fully confirmed our conclusions presented in Ref. \cite{bonitz_pre_comm}.
Finally, in their recent article  in Physica Scripta \cite{shukla_pscr} Shukla, Eliasson and Akbari-Moghanjoughi again massively criticized our analysis given in Ref. \cite{bonitz_pre_comm}.
% calling our paper ``dubious'' and our conclusions ``falsified'' and ``misleading'' (citations from the abstract of Ref. \cite{shukla_pscr}). 
% While we intend to ignore their inappropriate language, 
We cannot leave the additional statements of SEA in favor of QHD and against DFT 
as well as the new data for helium presented in Ref.~\cite{shukla_pscr} unanswered. In Sec.~\ref{s:arguments} we discuss these arguments in detail and show that they are incorrect and do not alter any of our conclusions presented in Ref.~\cite{bonitz_pre_comm}.

%\section{Linearized quantum hydrodynamics (LQHD)}\label{s:lqhd}

\section{The arguments of Shukla, Eliasson and Akbari-Moghanjoughi in Ref. \cite{shukla_pscr}}\label{s:arguments}
Let us first consider the arguments of SEA in support of the linearized QHD approach that was used in the original Letter \cite{shukla_prl}.
\begin{description}
 \item[i] The first objection of SEA against our analyis of the SE attractive potential is that (citation from Ref. \cite{shukla_pscr}, p. 3) the figures 1 and 2 of Ref. \cite{bonitz_pre_comm}
``are not compatible with our figures 1 and 2 here, because they have plotted $\alpha$ against different parameter, $r_s = r_0/a_B$ and
 $e\phi$ (eV) versus $r (a_B)$, respectively, which is different from
our $\alpha$ against $a_B/r_0$ and $\phi/k_sQ$ versus $k_sr$''. 

This is not correct. The corresponding figures of $\alpha$ and the potential in the two papers are trivially translated into each other by properly rescaling the respective axes. Furthermore, they give an incorrect definition of the parameter $r_s$ used in Ref.~\cite{bonitz_pre_comm} using, instead, the parameter $r^{\rm SEA}_s$, see the definitions (\ref{eq:rs}) and (\ref{eq:rs_sea}).
 \item[ii] SEA underline explicitly (p. 3): ``It is important to note that, the correct criterion for the QHD equation validity [9, 10] is $n_0 \lambda_B^3 \ge 1$
and not $\hbar\omega_{\rm pe}/k_BT_F < 1$''. 

This is wrong. This statement would mean that QHD is always valid whenever the electrons are quantum degenerate, see above. [Note that, since QHD is a theory for $T=0$, $\lambda_B \to \infty$, according to Eq. (\ref{eq:lambda_b}), and QHD would be always correct, which obviously is impossible.]
However, the condition $n_0 \lambda_B^3 \ge 1$ says nothing about the coupling strength of the plasma. QHD, in fact, is a theory for the (nearly) ideal electron gas, i.e. it does not apply 
to a plasma with $r_s > 1$ or $\hbar\omega_{\rm pe}/k_BT_F > 1$ \cite{bonitz_pre_comm,bonitz_pre_reply,vladimirov}. 

In that context it is interesting what the same authors write in different papers. In the beginning of their Letter \cite{shukla_prl} Shukla and Eliasson themselves formulate the condition $\hbar\omega_{\rm pe}/k_BT_F < 1$, but in the following ignore it and present data for $\hbar\omega_{\rm pe}/k_BT_F > 1$. Furthermore, in their answer to our analysis given in Ref. \cite{shukla_answer} the same authors state that it would be obvious that QHD is valid only for an ideal quantum plasma, i.e. for $r_s \to 0$ and $\Gamma_q \to 0$.
 \item[iii] On p. 3 SEA continue: ``In fact, the SE model for the screening of a test charge [2] in quantum plasmas is analytic and exact, since
it includes the most important quantum electron interference effect''.

This is wrong. Obviously, the QHD is not an exact theory, due to the existing limitations \cite{bonitz_pre_comm,bonitz_pre_reply,vladimirov}. Furthermore, QHD is a theory involving only the density (the absolute value of the electron wave function) and thus neglects the phase information. Therefore, it cannot describe coherence effects in principle \cite{bonitz_pre_reply}.
Statements about ``exactness'' of the SEA potential are repeated several times on pages 3 and 4. 
As a justification they write that this potential is computed ``directly from first principle Fourier transformation of Poison's equation without redundant assumptions''.
%
%This is a rather unusual understanding of a first principle theory.
%
%\item[v] On page 4 SEA again write about ``the exact analytical SE potential'', and a few lines later: ``it should be reiterated that
%Shukla and Eliasson [2] have calculated the exact expression for the short-range (of the order of a few aB) screening
%potential around a test ion charge ... discarding any artificial assumptions.''.
%This is again wrong for the reasons just mentioned.
%
 \item[iv] In the beginning of their paper SEA write: ``In their pioneering PRL paper, Shukla and Eliasson ... discovered a novel short-range attractive
force that can bring ions closer at atomic scales.''

This is wrong since QHD, as any hydrodynamic theory, uses an averaging procedure over scales of several inter-particle distances and, therefore cannot, by construction, resolve effects on the atomic scale \cite{bonitz_pre_comm,bonitz_pre_reply,vladimirov}.
\item[v] On page 2 SEA provide another argument in support of the QHD based on its description of electronic plasma oscillations (EPOs): ``the validity
of the linearized QHD equations and their applications have been put on a firm footing, because the frequency
spectra of EPOs have indeed been experimentally verified by Watanabe [15] in metals and by Glenzer et al [16] in warm
dense matter''.

The fact that QHD yields the correct long-wavelength limit of the plasma oscillation spectrum of an ideal electron gas, 
in agreement with the RPA result \cite{klimontovich,bohm_pines} (see above)
% and recent experiments, is certainly a necessary condition for the validity of QHD but by no means a proof of its general correctness or of 
%its applicability to dense hydrogen. In particular, this agreement 
is irrelevant for the small scale behavior (short wavelengths) where the SE potential is predicted to be attractive.
\end{description}
In summary, the above statements of SEA about the QHD are invalid. 
%and its validity provide no facts proving the validity of the SE attractive potential. 
Furthermore, SEA did not present any facts against the arguments of Ref. \cite{bonitz_pre_comm} that rule out the attractive potential for dense hydrogen, cf. points 1.-5. in Sec.~\ref{s:dispute}. 
%Interestingly, SEA did indirectly respond to one of our arguments -- about the incompability of the minimum position of the SE potential with the interparticle distance (point 2 above) by providing 
Instead they presented new data for another material -- helium. 
%Apparantly no such value exists for hydrogen, in contrast to the original predictions of Shukla and Eliasson \cite{shukla_prl}.
%Yet the new data for helium do not withstand any criticism, 
However, even without knowing any details of the authors' calculations or information about the type of ``$^4$He-plasma'' they have investigated, a simple estimate shows that the minimum position of the potential provided in Ref. \cite{shukla_pscr} corresponds to 
a value of the Brueckner parameter, $r_s \approx 1.43$. At such a density and $T=0$ helium is practically completely neutral, e.g. \cite{schlanges}, and no helium ions and free electrons exist, nor is there an attractive ion-ion interaction. Clearly, this parameter region corresponds to the strongly coupled plasma state where QHD is invalid (see item ii. above).

Let us now consider the arguments of SEA against the validity of the density functional simulations of Ref. \cite{bonitz_pre_comm}. We use italics to underline the key points in quotes from Ref.~\cite{shukla_pscr}.
\begin{description}
 \item[a] On p. 3 SEA write: ``Henceforth, it is concluded that the standard DFT simulation results of BPS ... have to be discarded {\em since they are incompatible with the newly discovered SE attractive force} that brings ions closer at atomic scales to form a high density superconducting plasma state with temperatures of the order of a few electron volts.''.
%In short: DFT must be wrong because it contradicts QHD. Furthermore, we learn that the SE attractive potential gives rise to superconductivity. Unfortunately, SEA do not uncover the secret about the physical mechanism.
%
% \item[b] 
On p. 3 SEA continue: ``The results of BPS [1] should be revised to include the energy associated with the quantum
recoil effect in the DFT Hamiltonian in order {\em to obtain a consistent output, similar to that of Shukla and Eliasson} ...'' 
%Thus, SEA suggest to modify the DFT simulations such that they will agree with QHD. This is, of course, in striking contrast to the {\em ab initio} 
%concept of DFT.
%
% \item[c] 
SEA continue by noting that ``{\em the Bohm potential}, $V_B = {\hbar}^2/2m_e \Delta \sqrt{n(r)}/\sqrt{n(r)}$ {\em is missing in the DFT Kohn--Sham
Hamiltonian}.''  On p. 3 SEA write that electron ``tunneling through the quantum Bohm potential which is absent ...
in the density functional theory (DFT) used for simulations of BPS ..., completely ignoring the quantum electron tunneling
effect''.

These statements are  not correct. DFT starts from the full many-electron hamiltonian. 
The strength of the {\it ab initio} approach of Hohenberg and Kohn \cite{hk_DFT} and Kohn and Sham \cite{ks_DFT} is to map 
the interacting $N$-body quantum problem onto a system of non-interacting particles in an effective potential.
For practical calculations, just a single approximation is required: 
the form of the exchange-correlation potential $V_{XC}$, see \cite{bonitz_pre_comm} and references therein.
Therefore, no additional potentials, such as 
the Bohm potential, can be artificially included in this scheme. Note that the Bohm potential is derived in QHD from the quantum kinetic 
energy (Laplacian of the wave function) which is fully included in DFT. 
In particular, DFT fully describes electron tunneling effects \cite{bonitz_pre_reply}.
 \item[b] On p. 3 SEA write about our simulations: ``...their DFT theory, which uses the simplified Thomas–Fermi ionic potential in ground-state
density functional, is unable to capture the essential physics of quantum density spreading effects at atomic scales in quantum
plasmas.'' 
Further, on p. 3 SEA write that the ``ion potential calculation follows ... from the pseudopotential approximation ...which takes only the effect of valence electrons...''.
%SEA continue: ``This method is based on {\em many approximations} ..., which limits its general applicability''.

These statements are not correct. The DFT simulations of Ref. \cite{bonitz_pre_comm} have no relation to Thomas-Fermi ionic potentials, and no such information was given in Ref. \cite{bonitz_pre_comm}. There it has also been clearly stated that no pseudopotential has been used but the exact Coulomb potential. Further, as mentioned above, DFT fully captures quantum diffraction effects (``quantum density spreading''), see also \cite{bonitz_pre_reply}. As mentioned above, DFT includes a single approximation, the choice of the potential $V_{XC}$.
\item[c] SEA write that DFT is ``based on the work by Hohenberg and Kohn ..., where it has been
{\em assumed} that the ground-state energy of electron is a unique property of the electron number density''.
SEA continue on p. 3: ``DFT simulations also rely on the Kohn and Sham’s {\em assumption} ... that the energy functional
can be recast by using electronic orbital as $E_{KS}(\{\Psi_i\})$...''.

%These statements contradict well-known facts: 
We point out that Hohenberg and Kohn and Kohn and Sham have proven theorems, not mere assumptions, which constitute the basis of DFT.
\item[d] SEA write on p. 4 about computational aspects: ``DFT simulations are usually carried out using the periodic boundary conditions, which are
most useful for crystalline matter band structure calculations and are hence irrelevant for non-crystalline plasma-like
medium.'' SEA write: ``In a simulation of single-electron plane wave
calculations, yet another approximation is involved which is called the plane-wave cut-off approximation ...
By using symmetry considerations, the computational time is greatly saved in such
simulations. However, for the case of asymmetric boundless electron-ion plasmas the simplified DFT calculation becomes
uninformative.''

This is not correct. In contrast, real DFT simulations achieve high accuracy for macroscopic (unbounded plasmas), 
e.g. \cite{bonitz_pre_comm, militzer12}. Furthermore, non-crystalline systems can be simulated within supercells using perdiodic boundary conditions (not only) in DFT.
\item[e] On p. 5 SEA write about DFT: ``nonuniform electron-distribution ...
and ion/electron collisions ... are ignored.''

This is not correct. DFT accurately describes nonuniform electron distributions in atoms, molecules, condensed matter systems or plasmas 
as is demonstrated in many papers and text books. The treatment of electron-electron interaction effects is determined by the 
quality of the exchange-correlation functional.
\item[f] On p. 5, SEA ``emphasize that BPS ... overlooked the free-electron assumption for their simple hydrogen plasma,
since DFT theory does not resolve such boundary as the electron degeneracy.''

This statement makes no sense. Dense hydrogen is not describable as a free electron gas nor is such an assumption included in DFT -- in contrast to QHD. 
Moreover, electron degeneracy is fully included in DFT. Note that, in order to adequately compare with the linearized QHD results of Ref.~\cite{shukla_prl},
the present authors in Ref.~\cite{bonitz_pre_comm} performed DFT simulations for protons embedded in jellium although this model is not regarded as the ultimate theoretical approach to dense hydrogen, as was clearly stated~\cite{bonitz_pre_comm}.
\item[g] On p. 4 SEA claim that the present authors ``{\em misled the quantum plasma physics community} by mentioning that Almbladh
et al. \cite{almbladh76} and Bonev and Ashcroft \cite{ashcroft01} already have treated the problem of screening of a proton in an electrons gas,
including relevant quantum forces that are of physical interest''. [The numbers used to refer to the references have been adjusted to correspond to those relevant for and used in the present manuscript.]

Readers are encouraged to look up these references and convince themselves that 
already more than 35 years ago accurate DFT simulations of the proton potential have been performed \cite{almbladh76}
 and clearly demonstrated the crucial importance of nonlinear effects for this problem.
\end{description}

Summarizing the discussion of SEA on DFT we conclude that Shukla et al. draw a distorted picture of this well established and highly successful quantum simulation method. SEA incorrectly assert that DFT neglects basic quantum effects, such as electron tunneling and diffraction. 
At the end of their analysis of DFT SEA write (p. 5) that despite our ``claim on the exactness of their DFT-based simulation results, such
approaches are under extensive improvements''.
%
%We reject this statement. In our paper \cite{bonitz_pre_comm} we clearly stated the approximate nature of DFT the quality of which is determined (exclusively) by the exchange-correlation functional, and we presented all necessary details about the choice we made together with convergence tests. Nowhere in our paper is there a statement that DFT is ``exact''.
In fact, as many theoretical and computational methods for nonideal quantum systems, DFT is under active development. Yet this does not mean 
%-- as SEA are trying to suggest -- 
that the presently available DFT simulations are unreliable. For further details on DFT the reader is referred to the references listed in \cite{bonitz_pre_comm}.

In conclusion, we have given a detailed analysis of the arguments of SEA presented in their Physica Scripta paper \cite{shukla_pscr} providing a list of 
central statements. Our analysis revealed that, in this paper, there is not a single scientifically valid argument that would strengthen the validity of the LQHD result for the SE attractive potential, nor is there any valid argument that challenges our DFT results. Further, the new data presented by SEA for helium do not support the SE potential or the formation of an ionic crystal.
Therefore, our conclusions presented in Refs. \cite{bonitz_pre_comm,bonitz_pre_reply} about the non-existence of the Shukla-Eliasson attractive potential of protons in dense hydrogen remain fully valid. 

This work is supported by the Deutsche Forschungsgemeinschaft via SFB-TR 24 project A5.
% and a grant for computing time at the North-German Supercomputing Alliance (HLRN). 

\appendix
\section{Summary of the SEAP potential}
The purpose of this abstract is to summarize the formulas leading to the Shukla-Eliasson (SE) attractive potential.
First we briefly summarize linearized quantum hydrodynamic theory (LQHD), including the particular approximations used by SE, and then list the predictions given by SE for dense hydrogen in Ref.~\cite{shukla_prl}. While in the main text we have used SI units, here we retain the notation and Gauss units used by SE in Ref.~\cite{shukla_prl}.

%LQHD is the linearized version of 
%quantum hydrodynamcs (QHD) \cite{manfredi-haas01, manfredi08}.\\[0.4cm]
%\subsection{QHD}
QHD \cite{manfredi-haas01, manfredi08} is obtained from the zero temperature classical hydrodynamic equations for the electron density $n$ and electron fluid velocity $\vec{\mathrm{u}}$, coupled to the electrostatic potential $\phi$ where a positive point charge $Q$ is assumed to be at ${\bf r}=0$):
%\begin{itemize}
% \item Continuity equation
\begin{eqnarray}
\frac{\partial n}{\partial t}+\nabla\cdot(n\vec{\mathrm{u}}) &=& 0,
\nonumber\\
%\item Momentum balance equation
m_*\left(\frac{\partial \vec{\mathrm{u}}}{\partial t} + \vec{\mathrm{u}} \cdot \nabla \vec{\mathrm{u}} \right) &=& e\nabla\phi-n^{-1}\nabla P + \nabla V_{xc} + \nabla V_B,
%\label{eq:dens_op}
\nonumber\\
%\item Poisson's equation
\nabla^2\phi &=& \frac{4\pi e}{\epsilon}(n-n_0)-4\pi Q\delta(\vec{r}).
\nonumber
\end{eqnarray}
Quantum effects are taken into account approximately, by adding the Bohm potential $V_B$
%\begin{itemize}
%  \item 
(which can be derived from the Schr\"{o}dinger equation and approximately accounts for quantum diffraction effects \cite{springer-buch}),
the pressure $P$ of the ideal Fermi gas at zero temperature (instead of the classical pressure) and an exchange-correlation potential $V_{xc}$ that 
was introduced in Ref. \cite{manfredi08} to account for exchange and correlation effects,
  \begin{eqnarray}
V_B &=& (\hbar^2/2m_*)(1/\sqrt{n})\nabla^2\sqrt{n},
\nonumber\\
P &=& (n_0m_*v_*^2/5)(n/n_0)^{5/3},
\nonumber\\   
V_{xc} &=&\begin{aligned}[t]&0.985(e^2/\epsilon)n^{1/3}\times\\
&\times[1+(0.034/a_Bn^{1/3})\ln(1+18.37a_Bn^{1/3})],\end{aligned}
\nonumber
\end{eqnarray}
with the definitions
$v_* = \hbar(3\pi^2)^{1/3}/m_*r_0$ and  $r_0 =n_0^{-1/3}$. Further, 
%a_B &= \epsilon\hbar^2/m_*e^2.
 $v_*$ is the electron Fermi velocity, $a_B$ the effective Bohr radius,
$m_*$  denotes the electron effective mass and $\epsilon$ the relative dielectric permeability of the material. In the calculations of Ref. \cite{shukla_prl} 
the authors used $m_*=m_e$ and $\epsilon=1$.

{\bf Linearization. LQHD.}
The equations of LQHD follow from the above nonlinear QHD equations in the linear approximation, where $n = n_0 + n_1$ and $|n_1| \ll n_0 $. If one further neglects dynamic effects, $\varepsilon(\vec{\mathrm{k}},\omega)=\varepsilon(\vec{\mathrm{k}},0)$, the electrostatic potential is given by \cite{shukla_prl}
\begin{equation}\label{a:phi}
\phi(\vec{r}) = \frac{Q}{2\pi^2}\int\frac{\exp(i\vec{\mathrm{k}}\cdot\vec{\mathrm{r}})}{k^2\varepsilon(\vec{\mathrm{k}})}d^3k
\end{equation}
with the result for the inverse dielectric function
 \begin{equation}
\frac{1}{\varepsilon(k)} = \frac{(k^2/k_s^2)+\alpha k^4/k_s^4}{1+(k^2/k_s^2)+\alpha k^4/k_s^4},
\end{equation}
\vspace{-0.2cm}
with the definitions 
\begin{align}
\alpha &= \hbar^2\omega_{pe}^2/4m_*^2(v_*^2/3+v_{ex}^2)^2 \\
k_s &= \omega_{pe}/\sqrt{v_*^2/3+v_{ex}^2}, \\
v_{ex} &= \begin{aligned}[t] &({0.328e^2/m_*\epsilon r_0})^{1/2}\times\\&\times{[1+0.62/(1+18.36 a_B {n_0}^{1/3})]^{1/2}} \end{aligned}\\
k_r &= (k_s/\sqrt{4\alpha})(\sqrt{4\alpha}+1)^{1/2}
\nonumber\\
k_i &= (k_s/\sqrt{4\alpha})(\sqrt{4\alpha}-1)^{1/2}
\nonumber
\end{align}
and the plasma frequency $\omega_{pe} = \sqrt{4\pi n_0e^2/\epsilon m_*}$. The parameter $\alpha$ is shown in Fig.~\ref{fig:alpha}.

%\subsection{Predictions of LQHD}
{\bf Predictions of LQHD.}
While for $\alpha < 0.25$ the potential (\ref{a:phi}) is always positive, for $\alpha > 0.25$ it develops a negative (attractive) minimum. In the latter case it is given by 
\begin{equation}
\phi(\vec{r})=\frac{Q}{r}[\cos(k_i r)+b_*\sin(k_i r)]\exp(-k_r r),
\label{eqap:cos_pot}
\end{equation}
where $b_* =1/\sqrt{4\alpha-1} $.
The maximum value of $\alpha$ is approximately $0.64$. Inserting all parameters in the definition of $\alpha$, existence of a negative potential 
can be related to a finite density interval where $0.61 \le r_s \le 26.22$, that corresponds to moderate to strong coupling where QHD is not applicable, see Fig.~\ref{fig:alpha}. 
For weak coupling, $r_s \ll 1$, LQHD does not predict a negative potential.
\end{document}